\documentstyle[aps,prd,preprint]{revtex}

\begin{document}
\preprint{SNUTP 04-013}
\draft
\title{Localizing global hedgehogs on the brane}

\author{Inyong Cho\footnote{Electronic address:
iycho@phya.snu.ac.kr}}
\address{Center for Theoretical Physics, School of Physics,
Seoul National University, Seoul 151-747, Korea}

\date{\today}

\maketitle

\begin{abstract}
We investigate the localization of 4D topological global defects on the brane
embedded in 5D.
The defects are induced by 5D scalar fields with a symmetry-breaking potential.
Taking an ansatz which separates the scalar field into the 4D and the extra-D part,
we find that the static-hedgehog configuration is accomplished and
the defects are formed
only in the $AdS_4/AdS_5$ background.
In the extra dimension, the localization amplitude for the 4D defects
is high where the warp factor is high.
\end{abstract}

\vspace{24pt}
\pacs{PACS numbers: 98.80.Cq, 11.10.Kk, 04.50.+h}

\section{Introduction}
Since the large extra dimension scenario was introduced~\cite{ADD,RS},
a lot of attention has been paid to localizing matter fields
on the brane worlds (see, for example, Refs.~\cite{Shifman,Bajc,Dubovsky}).
The matter fields one can consider in the field theory are the scalar field,
gauge field, and spinor field.
For two-brane models, the localization (normalization) of such fields is
always possible by an appropriate compactification of the extra dimensions
bounded by the branes.
For one-brane models, however, the success of localization depends on the background
bulk geometry and the field configuration in the bulk.

As an example in 5D, for free-boson fields of spin zero and one, the action is
\begin{equation}
S =\int d^4x dy \sqrt{-g} \left( -{1\over 2} \partial_A \Phi \partial^A\Phi
-{1\over 4} F_{AB}F^{AB}\right)\,,
\end{equation}
where the indices run over the whole five dimensions,
and the field strength of the spin one
field is determined by the gauge field, $F_{AB} = \partial_A W_B -\partial_B W_A$.
In order to consider localization,
one usually applies a ``separation ansatz'' on the fields,
\begin{eqnarray}
\Phi (x^\mu,y) &=& \phi(x^\mu) p(y)\,,\\
W_\alpha (x^\mu,y) &=& \omega_\alpha (x^\mu) q(y)\,,\quad W_y=0\,.
\end{eqnarray}
With a metric, for example,
\begin{equation}
ds^2 = B(y) \left( \overline{g}_{\mu\nu} dx^\mu dx^\nu +dy^2\right)\,,
\end{equation}
one can saturate the 4D part in the action as
\begin{eqnarray}
S &=& \int dy B^{3/2}(y) p^2(y) \int d^4x \sqrt{-\overline{g}}
\left( -{1\over 2}\overline{g}^{\mu\nu} \partial_\mu\phi\partial_\nu\phi \right)
+ ...\nonumber\\
{} &+& \int dy B^{1/2}(y) q^2(y) \int d^4x \sqrt{-\overline{g}}
\left( -{1\over 4}\overline{g}^{\mu\alpha}\overline{g}^{\nu\beta}
f_{\mu\nu} f_{\alpha\beta} \right)
+ ...\,,
\end{eqnarray}
where $f_{\mu\nu} = \partial_\mu \omega_\nu - \partial_\nu \omega_\mu$ is the usual
4D field strength.
The normalization of these 4D fields is thus determined by the warp geometry $B(y)$
and the transverse profile of the fields, $p(y)$ and $q(y)$.
The field is normalizable if the $y$-integration is finite.

In this paper, we are interested in the localization of 4D topological global defects
which are induced by 5D scalar-field multiplets $\Phi^a$, with a symmetry-breaking
potential $V(\Phi^a)$.
Being different from the ``usual'' scalar field,
a special ansatz for the scalar field, so called,
the ``hedgehog configuration'' is imposed
in order to realize topological defects; the scalar field
asymptotically approaches the vacuum-expectation value (the symmetry-breaking scale)
at large distances from the center while it remains in the unbroken-symmetry state
at the center~\cite{Vilenkin1}.
It is interesting to see if such extended bodies can be induced in 4D
by 5D scalar fields, and to see how they are localized in the extra dimension.
We shall impose this ansatz along the 4D part of the 5D scalar field, and investigate
the localization behavior by the transverse part of the scalar field.

As usual in the investigation of the localization of the other matter fields,
we do not consider self-gravity of the scalar field.
The background geometry is fixed.

We require the field to be ``separable'' into the 4D part and the transverse part.
Then as we observed above, the localization of 4D defects will be described by the
transverse part of $\Phi^a$ and the background geometry.

\section{Formation of 4D hedgehogs induced by 5D scalar fields}
We extend the usual 4D action for the scalar field with a symmetry-breaking potential,
which gives rise to topological defects.
The 5D action then reads
\begin{equation}
S=\int d^4xdy\sqrt{-g} \left[ -{1\over 2}\partial_A\Phi^a \partial^A \Phi^a
-{\lambda\over 4}(\Phi^a\Phi^a -\eta^2)^2\right]
-\int_i d^4x\sqrt{-h}\Sigma_i(\Phi^a)\,,
\label{eq=S1}
\end{equation}
where the spacetime index $A$ runs over 5D,
$a$ represents the field-space index,
and $\eta$ is the symmetry-breaking scale.
The boundary term appears when the model involves nonzero-tension branes located
at $y=y_i$.

We use a conformal nonfactorizable metric ansatz,
\begin{equation}
ds^2 = B(y) (\overline{g}_{\mu\nu}dx^\mu dx^\nu +dy^2)\,.
\end{equation}
We do not consider self-gravity of the scalar field, so the background geometry
[$B(y)$ and $\overline{g}_{\mu\nu}$] is given by introducing other matter fields such as
a bulk cosmological constant, or the brane tension.

For the scalar field, we apply a strictly ``separable'' ansatz as usual,
\begin{equation}
\Phi^a (x^\mu,y) = \phi^a (x^\mu) p(y)\,,
\end{equation}
and apply a ``static-hedgehog'' ansatz on the 4D part in order to realize topological
defects in 4D,
\begin{equation}
\phi^a (x^\mu) = f(r)\hat{\Omega}^a\,.
\end{equation}
Here, $\hat{\Omega}^a$ is the unit vector surfacing on a sphere $S^{a-1}$,
and $a=1,2,3$ for domain walls, cosmic strings,
and monopoles, respectively.
The static radial configuration $f(r)$ asymptotically approaches a constant, while
it takes $f(0)=0$.
This is a nontrivial configuration which gives rise to the hedgehog configuration.

With the above ans\"atze for the metric and the scalar field, the 4D part
of the action saturates
as we had seen in Introduction,
\begin{equation}
S = \int dy {\cal A}(y) \int d^4x \sqrt{-\overline{g}}
\left( -{1\over 2}\overline{g}^{\mu\nu} \partial_\mu\phi^a\partial_\nu\phi^a \right)
+ ...\,,
\label{eq=S2}
\end{equation}
where ${\cal A}(y) =B^{3/2}(y) p^2(y)$ is the normalization factor, and is interpreted
as the amplitude of finding the 4D configuration at the given location in the transverse direction.
This quantity ${\cal A}(y)$ determines the localization after the $y$-profile $p(y)$ is
obtained.

The field equation for $\Phi^a$ is given by
\begin{equation}
\nabla^A\partial_A \Phi^a = {\partial V(\Phi^a) \over \partial\Phi^a}
+{\sqrt{-h} \over \sqrt{-g}} {\partial\Sigma_i(\Phi^a) \over \partial\Phi^a}\delta(y-y_i)\,,
\end{equation}
where $h$ is the 4D metric density including $B(y)$.
This equation reduces to
\begin{equation}
{p''\over p} +{3 \over 2}{B' \over B}{p' \over p} +
{\overline{\Box}_{(4)} \phi^a (x^\mu) \over \phi^a (x^\mu)} = B\lambda
(p^2f^2-\eta^2) + B{\sqrt{-h}\over \sqrt{-g}}\left( {1\over
\Phi^a}{\partial\Sigma_i \over \partial\Phi^a}\right)
\delta(y-y_i)\,,
\label{eq=p1}
\end{equation}
where the prime denotes the derivative with respect to $y$.

Concerning the separation of variables, each term in the above equation is
a function of only $x^\mu$, or $y$ except the one $\lambda B(y)p^2(y)f^2(r)$.
The way to complete the separation is to make this term depend on only one variable.
First, consider $f(r) = \text{constant}$, which makes the term only $y$-dependent.
However, $f(r) =\text{constant}$ is a trivial solution which does not produce
the hedgehog configuration in 4D.
Therefore, we require a condition for the separation,
\begin{equation}
B(y)p^2(y) = \text{constant} \equiv c_1\,.
\label{eq=c1}
\end{equation}
Then the separation becomes complete, and Eq.~(\ref{eq=p1}) separates into two parts,
\begin{eqnarray}
{p''\over p}
+{3 \over 2}{B' \over B}{p' \over p} +
\lambda\eta^2B
-\sqrt{B}\left( {1\over \Phi^a}{\partial\Sigma_i
\over \partial\Phi^a}\right)\delta(y-y_i) &\equiv & m^2\,,
\label{eq=p}\\
{\overline{\Box}_{(4)} \phi^a (x^\mu) \over \phi^a (x^\mu)} -
\lambda c_1f^2(r) &=& -m^2\,,
\label{eq=f}
\end{eqnarray}
where $m^2$ is a constant.

First, note that the 4D equation~(\ref{eq=f}) is nothing but the usual 4D scalar-field equation
with a 4D effective symmetry-breaking potential,
\begin{equation}
\overline{V}(\phi^a) = {\overline{\lambda} \over 4}
(\phi^a\phi^a-\overline{\eta}^2)^2\,.
\label{eq=V4}
\end{equation}
We can identify
\begin{eqnarray}
\overline{\lambda} &=& \lambda c_1\,,\\
m^2 &=& \overline{\lambda}\overline{\eta}^2\,.
\label{eq=m2}
\end{eqnarray}
The latter relation~(\ref{eq=m2}) states that
$m$ defined in Eqs.~(\ref{eq=p}) and (\ref{eq=f})
behaves like the mass of the 4D perturbed-scalar field where the symmetry is broken.
We will see later that the sign of $m^2$ plays an important role in defect formation.

Second, we can cast the extra-D equation~(\ref{eq=p}) in a differential equation only for $B$ by using
the separation condition $Bp^2 = c_1$ [Eq.~(\ref{eq=c1})].
Then the equation leads to
\begin{equation}
B''+2m^2B -2\lambda\eta^2B^2 + 2B^{3/2}\left( {1\over \Phi^a}{\partial\Sigma_i \over
\partial\Phi^a} \right)\delta(y-y_i) =0\,.
\label{eq=B}
\end{equation}
As a result, the separation condition converted the scalar-field equation to an Einstein
equation.

From the beginning we have ignored self-gravity of the scalar field, and
have assumed that the background geometry is given and fixed
by some other matter fields.
Such a given background, however, does not necessarily satisfy the above
Einstein equation~(\ref{eq=B}).
Therefore, what is remaining is to investigate which backgrounds have Eq.~(\ref{eq=B}) be
consistent with their Einstein equations.

Let us analyze Eq.~(\ref{eq=B}).
The last term is the boundary term which arises when there exist nonzero-tension branes.
The brane tension $\sigma_i$ is related as
\begin{equation}
\kappa^2\sigma_i = 3\left[ {1\over \Phi^a}{\partial\Sigma_i(\Phi^a) \over \partial \Phi^a}\right]_{y=y_i}\,,
\label{eq=sigmai}
\end{equation}
where $\kappa^2 =8\pi G_5$.

The quadratic term in $B$ arises when there exists a bulk cosmological constant $\Lambda_5$
which is to be related to the parameters from Eq.~(\ref{eq=B}),
\begin{equation}
\Lambda_5=-2\lambda\eta^2\,.
\end{equation}
The linear term corresponds to the 4D curvature term, and the parameter is related as
\begin{equation}
\overline{R}_{(4)} =-6m^2\,.
\end{equation}

If the background-matter fields are provided
in such a way to satisfy the above relations,
Eq.~(\ref{eq=B}) is completely satisfied.

From $\Lambda_5 =-2\lambda\eta^2$, $\Lambda_5$ can be either positive, or negative
depending on the sign of $\lambda$.
If we keep $V(\Phi^a)$ in the conventional shape of the symmetry-breaking potential,
i.e., $\lambda >0$ and $\eta^2 >0$,
$\Lambda_5$ is negative.
Therefore, the bulk geometry is $AdS_5$.
$\Lambda_5$ is a free parameter, and its value is not restricted.

From $\overline{R}_{(4)} =-6m^2$, the 4D world-volume has a constant curvature.
There are several types of manifold of which curvature is constant.
First, we can consider the 4D Minkowski manifold of which curvature vanishes,
$\overline{R}_{(4)} = -6m^2 =0$.
Second, we can consider a manifold with a 4D cosmological constant,
and the curvature is $\overline{R}_{(4)} = -6m^2 = 4\Lambda_4$.
Therefore, the possible 4D manifold is one of $M_4$, $dS_4$, and $AdS_4$.

Let us examine these three 4D manifolds.
The 4D curvature is related with the mass parameter,
$m^2=\overline{\lambda}\overline{\eta}^2$ from Eq.~(\ref{eq=m2}).

(i) $M_4$: For the Minkowski manifold, $m^2=0$.
In the effective 4D potential $\overline{V}(\phi^a)$ in Eq.~(\ref{eq=V4}),
the mass term (quadratic term) is missing, and there exist only the constant term
and the quartic term of $\phi^a$.
Therefore, the nonzero vacuum-expectation value (VEV) of $\phi^a$ does not
develop, and there is no symmetry breaking along the world-volume direction.
Defects are not formed.

(ii) $dS_4$: For the de Sitter manifold,
$m^2=\overline{\lambda}\overline{\eta}^2 = -{2\over 3}\Lambda_4 <0$.
First, if $\overline{\lambda}<0$ while $\overline{\eta}^2 >0$,
the effective 4D potential $\overline{V}$ is an upside-down form of the
usual symmetry-breaking potential.
It is hard to expect that a stable hedgehog configuration is achieved
with this type of potential.
Second, if $\overline{\eta}^2 <0$ while $\overline{\lambda}>0$,
the potential has the absolute minimum
at $|\phi^a|=0$ with a nonzero-vacuum energy.\footnote{If we apply
a perturbation on the scalar field about the vacuum, $|\phi^a|=0$,
in this case, the perturbed-scalar field has
$(\text{mass})^2 = -\overline{\lambda}\overline{\eta}^2/2 >0$.
Therefore, $m$ identified as in Eq.~(\ref{eq=m2}) does not exactly
represent the mass of the perturbed-scalar field in all cases.
We had better interpret Eq.~(\ref{eq=m2}) as just a relation between the parameters.}
There does not develop a nonzero VEV, and
the situation is similar to $M_4$. No defects are formed.

(iii) $AdS_4$: For the Anti-de Sitter manifold,
now the potential $\overline{V}$ has
the shape of the usual symmetry-breaking potential.
The nonzero VEV, $|\phi^a| = \overline{\eta}$,
develops and the defects are formed.

As a result, in generating the hedgehog configuration along the 4D world-volume
with the 5D scalar field and the symmetry-breaking potential,
the only possible manifold is $AdS_4/AdS_5$.

We have assumed that self-gravity of the scalar field is negligible,
and the background geometry is given by other matter fields.
In this case, the background is $AdS_4/AdS_5$ and it is generated by
$\Lambda_5$ and $\Lambda_4$.\footnote{In a general sense, $\Lambda_4$ is
not regarded as a source of the curvature.
Instead, the brane tension is a source, and
$\Lambda_4$ is an induced quantity by this tension and $\Lambda_5$ by matching
boundary conditions.
Here, however, we are not going to be bothered by such a discretion.}
Meanwhile, as we observed above,
these quantities are related with the parameters of the scalar field by
$\Lambda_5 = -2\lambda\eta^2$ and
$\Lambda_4= -{3\over 2}\overline{\lambda}\overline{\eta}^2$.
Then one may wonder how consistently we can keep
the parameters $\Lambda_5$ and $\Lambda_4$ as the source of curvature,
while we are ignoring the corresponding scalar-field parameters.
The clue is that the scalar-field contribution to the curvature
(to the energy-momentum tensor, in other words)
is suppressed by
the Planck-mass scale, so the condition,
$\eta \ll \mbox{Planck-sacle}$, is sufficient to
ignore its effect,
and the same reasoning is true  for $\Lambda_4$ and the 4D parameters.

\section{Review of $AdS_4/AdS_5$}
In the previous section, we observed that the 4D hedgehog configuration
is realized in the $AdS_4/AdS_5$ background.
In this section, we review the localization of gravity
in this background, which was investigated by
Karch and Randall~\cite{Karch},\footnote{Localization of the ordinary-matter fields
in this set-up (one-brane model) was investigated in Ref.~\cite{Oda}.}
and Kogan, {\it et. al.}~\cite{Kogan}.
We shall fucus on the two-brane model later on.

The Einstein-Hilbert action for gravity is
\begin{equation}
S_g = \int d^4xdy \sqrt{-g} \left[ {R-2\Lambda_5\over 2\kappa^2}
-{\sqrt{-h}\over \sqrt{-g}}\sigma_i\delta(y-y_i) \right]\,.
\end{equation}
With the metric ansatz,
\begin{equation}
ds^2 = B(y) (\overline{g}_{\mu\nu}^{AdS} dx^\mu dx^\nu +dy^2)\,,
\label{eq=metricBG}
\end{equation}
the solution to Einstein equations is given by
\begin{equation}
B(y) = 2{\Lambda_4\over \Lambda_5}\sec^2\left[ \sqrt{-{\Lambda_4\over 3}}(|y|-y_0)\right]\,,
\end{equation}
where $y_0$ is an integration constant which can be fixed if we apply a
normalization condition on $B$ [e.g., $B(0)=1$].
However, we will not do this, but leave $y_0$ free.

The 4D cosmological constant $\Lambda_4$ is determined by $\Lambda_5$ and
the brane tension,
\begin{eqnarray}
\sigma_I &=& {\sqrt{-6\Lambda_5} \over \kappa^2}
\sin\left(\sqrt{-{\Lambda_4\over 3}} y_0\right)\,,\\
\sigma_{II} &=& {\sqrt{-6\Lambda_5} \over \kappa^2}
\sin\left[\sqrt{-{\Lambda_4\over 3}} (y_*-y_0)\right]\,,
\end{eqnarray}
where brane I (II) is located at $y=0$ ($y=y_*$).

The warp factor $B(y)$ is a periodic function which diverges at
\begin{equation}
\sqrt{-{\Lambda_4 \over 3}}(|y|-y_0) = \left(n+{1\over 2} \right) \pi\,.
\end{equation}
In order to avoid the difficulties coming from the divergence,
we compactify the bulk to contain only the regular region
bounded by the branes.
Without loss of generality, we require $y_0>0$.
The warp factor $B(y)$ falls as $y$ increases from brane I at $y=0$,
and reaches the minimum at $y=y_0$, and then increases.
The tension of brane I is positive ($\sigma_I>0$),
and the tension of brane II changes its sign depending on the the brane location,
$\sigma_{II} >=< 0$ for $y_* >=< y_0$.
As was indicated in the previous section,
these tensions are related with the boundary terms
in the scalar-field action~(\ref{eq=S1})
by the relation~(\ref{eq=sigmai}), for consistency.

To investigate the localization of gravitons, introduce a perturbation to the
background metric,
\begin{equation}
ds^2 = g^{(bg)}_{AB}dx^Adx^B +h_{\mu\nu}dx^\mu dx^\nu\,,
\end{equation}
where $g^{(bg)}_{AB}$ represents the background metric~(\ref{eq=metricBG}),
and
\begin{equation}
h_{\mu\nu}(x^\mu,y)= h(y)\hat{e}_{\mu\nu}(x^\mu)\,.
\end{equation}
From the field equations,
\begin{eqnarray}
\Box_{(bg)}h_{MN} &+& 2R^{(bg)}_{MANB}h^{AB} = 0\,,\\
\overline{\Box}_{(4)} \hat{e}_{\mu\nu}(x^\mu) &=&
\left(m_g^2 +{2\over 3}\Lambda_4 \right)\hat{e}_{\mu\nu}(x^\mu)\,,
\end{eqnarray}
and by rescaling $\hat{h}(y) = B^{-1/4}h(y)$,
we obtain the nonrelativistic Schr\"odinger-type equation for the graviton,
\begin{equation}
\left[ -{1\over 2}{d^2 \over dy^2} + U(y)\right] \hat{h}(y) =
{m_g^2\over 2} \hat{h}(y)\,,
\end{equation}
where
\begin{eqnarray}
U(y) &=& -{\Lambda_4\over 3} \left\{ -{9\over 8}
+{15\over 8}\sec^2\left[\sqrt{-{\Lambda_4\over 3}}(|y|-y_0) \right]\right\}
+\tilde{\Sigma}_I\delta (y)
+\tilde{\Sigma}_{II}\delta (y-y_*)\,,\\
\tilde{\Sigma}_I &=& {3\over 2}\sqrt{-{\Lambda_4\over 3}}
\tan\left(\sqrt{-{\Lambda_4\over 3}}y_0 \right)\,,\\
\tilde{\Sigma}_{II} &=& {3\over 2}\sqrt{-{\Lambda_4\over 3}}
\tan\left[\sqrt{-{\Lambda_4\over 3}}(y_*-y_0) \right]\,.
\end{eqnarray}
The zero-mode solution ($m_g^2 = 0$) to this equation is given by
\begin{equation}
\hat{h}_0(y) = a_1 \sec^{3/2} \left[\sqrt{-{\Lambda_4\over 3}}(|y|-y_0) \right]\,,
\label{eq=h0}
\end{equation}
where $a_1$ is the normalization constant,
and the massive-mode solutions are described the hypergeometric function.

The zero-mode~(\ref{eq=h0}) is again a $sec$-function which diverges where
the warp factor $B$ diverges.
Therefore, in the one-brane model which bounds to the singular points,
the zero-mode is not normalizable.
This zero-mode does not recover 4D gravity~\cite{Karch}.
However, in the two-brane model, it is manifestly normalizable
by placing the second brane at an appropriate position~\cite{Kogan}.

In addition to the zero-mode, the graviton has an almost massless mode
which is the first-excited massive mode, $\hat{h}_1$.
In the one brane model of Ref.~\cite{Karch},
this ultra-light mode was responsible for 4D gravity
while the zero mode is absent (because it is not normalizable).
In the two brane model of Ref.~\cite{Kogan},
this mode contributes to 4D gravity almost equally
to the normalizable zero-mode.
In the latter, authors placed the second brane at the symmetric point
about the minimum of $B$, i.e., at $y=y_*=2y_0$.
And, they also discovered that 4D gravity remains in the same picture
even when the second brane is displaced very far from
the symmetric point, $y_* \gg 2y_0$.

In the next section, we shall discuss the hierarchies
in the two-brane set-up.
We shall place the second brane very far from the minimum of $B$, i.e.,
$y_*\gg y_0$ in order to achieve proper hierarchies of physical quantities
between the two branes.
With this brane separation, we shall discuss the localization
of the 4D defects by looking at the amplitude ${\cal A}(y)$.
Meanwhile, gravity will be localized in a similar manner
to the symmetric case $y_*=2y_0$
as was stated above, so we will not be concerned any further.

\section{Hierarchies and defect localization}
In this section, we discuss the hierarchies of various physical quantities
in the literature, and the localization amplitude of defects.
We assume a two-brane model where the second brane is displaced far from the
symmetric point.
For the symmetric case ($y_* = 2y_0$) in Ref.~\cite{Kogan},
there is no hierarchy between the two branes since the warp factor $B$ (which controls
the hierarchy) is the same at the two branes.

\subsection{Particle mass}
The hierarchy is achieved as the warp factor flows along the transverse direction.
The mass hierarchy of Higgs-type particles is explicitly described as (for derivation, readers
see Ref.~\cite{RS})
\begin{equation}
v_{eff} \propto \sqrt{B(y=y_i)}\,,
\end{equation}
where $v_{eff}$ is the mass scale and $y_i$ is the location of the brane.
Therefore, the particle-mass hierarchy between the two branes is given by
\begin{equation}
{v_{eff} (y=0)\over v_{eff}(y=y_*)} = \left[{B(0) \over B(y_*)}\right]^{1/2} \equiv \mu\,.
\end{equation}
For example, to achieve the TeV/Planck-scale hierarchy, $\mu \approx 10^{-15}$.
In this work, however, we will not restrict the absolute scale of hierarchies
between the branes. Instead, we shall discuss qualitatively how physical properties
flow along the extra dimension, and compare their relative picture between the branes.

\subsection{$G_4$}
The 4D gravitational constant at $y=y_i$ is given usually by the zero-mode graviton,
\begin{equation}
G_4(y=y_i) = G_5 |\hat{h}_0(y=y_i)|^2/e_l \propto B^{3/2}(y_i)\,,
\end{equation}
where the length unit $e_l$ was included to match the dimension.
In $AdS_4/AdS_5$, the ultra-light massive mode $\hat{h}_1$ will contribute to $G_4$
almost equally, but the above proportionality will not be altered much.
Therefore,
\begin{equation}
{G_4(0) \over G_4(y_*)} = \left[ {B(0) \over B(y_*)} \right]^{3/2} = \mu^3\,.
\end{equation}
The flowing picture is similar to that of the particle mass with only a different power.

\subsection{Effective 4D cosmological constant}
Now let us discuss the effective 4D cosmological constant on the brane.
For $(A)dS_4$ branes, the 4D metric at $y=y_i$ is given by
\begin{eqnarray}
ds_4^2(y=y_i) &=& B(y) \left[\overline{g}_{\mu\nu}^{(A)dS} dx^\mu dx^\nu \right]\\
{}&=& B(y_i) \left( 1+ {\Lambda_4\over 12}l^2\right)^{-2} (-dt^2 + \gamma_{ij}dx_i dx^j)\,.
\end{eqnarray}
where $l^2$ is the Lorentzian-length element,\footnote{For cartesian coordinates $(x_1,x_2,x_3)$,
$l^2 = -t^2 + x_1^2 +x_2^2 +x_3^2$, for cylindrical coordinates $(r,\theta,z)$, $l^2 = -t^2 +r^2+z^2$,
and for spherical coordinates $(r,\theta,\varphi)$, $l^2=-t^2+r^2$.}
and $\gamma_{ij}dx^idx^j$ is the 3D flat metric.
Absorb $B(y_i)$ by rescaling the coordinates,\footnote{The coordinates are
rescaled by $X^\mu = \sqrt{B(y_i)} x^\mu$
(the angular coordinates are not rescaled, but remain unchanged),
thereby $L=l\sqrt{B(y_i)}$.}
then the metric leads to
\begin{equation}
ds_4^2(y=y_i) = \left[ 1+{\Lambda_4 \over 12B(y_i)}L^2 \right]^{-2}(-dT^2 + \Gamma_{ij}dX^idX^j)\,,
\end{equation}
where $\Gamma_{ij}dX^idX^j$ is again the 3D flat metric
in the rescaled coordinates.
In this new coordinates, the effective 4D cosmological constant is given by
\begin{equation}
\Lambda_4^{eff}(y=y_i) = {\Lambda_4 \over B(y_i)}\,.
\end{equation}
This $\Lambda_4^{eff}$ is normalized if we normalize $B(y)$.
For example, if we normalize as $B(0)=1$, then $\Lambda_4^{eff}(0)=\Lambda_4$.
However, since we are interested only in the relative scales between the two branes,
we will not do such a normalization.

The resulting effective vacuum-energy density can be obtained by
\begin{equation}
\rho^{eff}_{\Lambda_4}(y=y_i) = {\Lambda_4^{eff}(y_i) \over 8\pi G_4(y_i)}
\propto {\Lambda_4 \over B^{5/2}(y_i)}\,,
\end{equation}
and we get the ratio,
\begin{equation}
{\rho^{eff}_{\Lambda_4}(0) \over \rho^{eff}_{\Lambda_4}(y_*)} =
\left[ {B(0) \over B(y_*)} \right]^{-5/2} = \mu^{-5}\,.
\end{equation}
The effective vacuum-energy density as well as the effective 4D cosmological constant
flows along the the extra dimension in the opposite way to the other physical
quantities discussed before.
Their scale is small where the warp factor is large.

Wherever our universe is located in the extra dimension, the effective 4D cosmological
constant should be bounded by some limit.
The observational data show that the recent accelerating expansion of our universe
prefers a positive cosmological constant.
However, milder restriction on the cosmological constant
is given by the Weinberg window~\cite{Weinberg},
\begin{equation}
-10^{-120} M_{Pl}^4 < \rho^{eff}_{\Lambda_4} < 10^{-118} M_{Pl}^4\,,
\end{equation}
which admits negative values, but very small.

\subsection{${\cal A}(y)$}
Finally let us discuss the localization amplitude ${\cal A}(y)$ of the 4D defects.
${\cal A}$ was defined from the normalization~(\ref{eq=S2})
of the scalar field along the extra dimension.
Imposing the separation condition~(\ref{eq=c1}), it becomes
\begin{equation}
{\cal A}(y) = B^{3/2}(y)p^2(y) = c_1\sqrt{B(y)}\,.
\label{eq=A}
\end{equation}
The ratio between the two branes becomes
\begin{equation}
{{\cal A}(0) \over {\cal A}(y_*)} =
\left[ B(0) \over B(y_*) \right]^{1/2} =\mu\,.
\end{equation}
The flow-pattern of the amplitude is the same with that of the particle mass $v_{eff}$.

We may interpret the amplitude ${\cal A}(y)$ as the probability density
to find 4D defects at a given location $y$.
Then, from the above result
the defects can be more likely localized where the warp factor is large.

Suppose that our universe is located on brane II where the warp factor is
large. Compared with brane I,
the effective 4D cosmological constant can be relatively very small,
while the localization amplitude of defects is  very high.
However, in this case, gravity becomes strong compared with that on brane I.

In this work, we have not been very interested in the absolute values of
the physical quantities ($G_4$, $\Lambda_4^{eff}$, etc.),
since we have been concerned mainly on the qualitative picture
of the hierarchy and the defect localization.
Therefore, we do not need to evaluate the values of the constants involved
with the model such as $y_0$, $y_*$, and $a_1$,
but let us complete this section with the derivation of $c_1$, the constant
appearing in the separation condition.
It is evaluated from the normalization,
\begin{equation}
N=\int^{y_*}_{-y_*} {\cal A}(y) dy = c_1\int^{y_*}_{-y_*}\sqrt{B(y)} dy =1\,,
\end{equation}
where we used Eq.~(\ref{eq=A}).
After integration, we get
\begin{equation}
c_1^{-1} = \sqrt{-{6\over \Lambda_5}}\ln\left[ {(1+\sin Y_*)\cos Y_0 \over
(1-\sin Y_0)\cos Y_*} \right]\,,
\end{equation}
where we defined, $Y\equiv y\sqrt{-\Lambda_4/3}$.

\section{Conclusions}
In this paper, we have investigated the localization of 4D topological global defects
induced by 5D scalar fields.
We adopted an ansatz for the scalar field $\Phi^a$,
which strictly ``separates'' $\Phi^a$ into
the 4D part $\phi^a(x^\mu)$ and the extra-D part $p(y)$.
The action involved the usual kinetic and symmetry-breaking potential terms,
which are covariantly extended to higher dimensions.
We also assumed a static ansatz for the fields.

We ignored self-gravity of the scalar field, and assumed that
the background geometry is given by some other matter fields such as $\Lambda_5$.
This is guaranteed if the scale of scalar-field parameters is small compared with the
corresponding Planck scale (for example, $\eta \ll \mbox{5D Planck scale}$, etc.).

Starting with the separation ansatz for the scalar field,
we found that
the scalar-field equation requires a specific condition
to make the equation separable;
the warp factor $B(y)$ and the $y$-profile of the scalar field, should satisfy
$B(y)p^2(y) =\text{constant}$.
The resulting two equations from the separation are the 4D equation and the extra-D equation.

The separation condition can convert the extra-D equation into an Einstein equation.
To be consistent with the resulting Einstein equation,
the background requires a bulk cosmological constant and
the constant 4D curvature.
The bulk cosmological constant is taken to be negative in order to keep the
symmetry-breaking potential in a conventional shape.

The 4D equation is nothing but the usual scalar-field equation for
defects.
The solutions to this equation are the usual-defect solutions in the literature.
The mass parameter $m^2$ of the effective potential is controlled
by the 4D curvature $\overline{R}_{(4)}$.
Depending on this curvature, the realization of the hedgehog configuration
is determined.
For the zero-curvature Minkowski  and the positive-curvature
de Sitter 4D manifolds, the potential does not allow a nonzero vacuum-expectation
value of the 4D scalar field.
The hedgehog configuration is not achieved, and the defects are not formed in
the 4D world-volume.
For the negative-curvature anti-de Sitter 4D manifold, however,
the defects are possibly formed.
Therefore, the possible geometry to accomplish 4D topological defects in the 4D world
is $AdS_4/AdS_5$.

In a warped geometry, physical scales flow along the extra dimension.
We found that the localization amplitude of the defects is high where
the warp factor is high.
There, the magnitude of the effective 4D cosmological constant is
relatively very small, but gravity is strongly coupled.

All the results in this work rely crucially on the ``separation condition''.
If we lift the separation ansatz, we may be able to get the defect formation
consistently even in the Minkowski and the de Sitter manifold.
However, the techniques involved in getting the solutions must be very tough.

Including self-gravity of the scalar field seems also to make the problem
complicated. The separation ansatz in the self-gravitating system does not
look very feasible.

Any type of global defects share the results of this work.
The localization picture is the same for domain walls, cosmic strings, and
monopoles. However, it is not very plausible to assume that the domain walls
are static~\cite{Vilenkin2}.

\acknowledgements
This work was supported by the BK21 project of the Ministry of
Education and Human Resources Development, Korea.

\end{document}